\documentclass[aps,epsfigure,twocolumn,superscriptaddress,showkeys,nofootinbib]{revtex4-1}
\usepackage[colorlinks=true,linkcolor=blue,urlcolor=blue,citecolor=blue,pdfusetitle]{hyperref}
\usepackage[utf8]{inputenc}
\usepackage[english]{babel}
\usepackage{amsmath}
\usepackage[caption = false]{subfig}
\usepackage{graphicx,epstopdf}
\usepackage{blindtext}
\usepackage[table,xcdraw]{xcolor}
\usepackage{lipsum}
\usepackage{amsfonts}
\usepackage{bbm}
\usepackage{amssymb}
\usepackage{enumerate}
\usepackage{color}
\usepackage{latexsym}
\usepackage{comment}
\usepackage{times,txfonts}
\usepackage[normalem]{ulem}
\graphicspath{ {./figures/} }
\usepackage{subcaption}
\usepackage{amsmath}
\usepackage{algorithm}
\usepackage{algpseudocode}
\usepackage{tikz}
\usetikzlibrary{quantikz}
\usepackage{booktabs}
\usepackage{ragged2e} 

\begin{document}

\title{Simulating Work Extraction in a Dinuclear Quantum Battery Using a Variational Quantum Algorithm}

%%%%%% Affiliations %%%%%%
\author{Lucas Q. Galv\~{a}o}
\email{lqgalvao3@gmail.com}
\affiliation{Latin American Quantum Computing Center, SENAI CIMATEC, Salvador, Brasil.}
\affiliation{QuIIN - Quantum Industrial Innovation, EMBRAPII CIMATEC Competence Center in Quantum Technologies, SENAI CIMATEC, Av. Orlando Gomes 1845, Salvador, BA, Brazil, CEP 41650-010.}

\author{Ana Clara das Neves}
\email{ana.neves@ufabc.edu.br}
\affiliation{Universidade Federal do ABC, Av. dos Estados, 5001 - Bangú, Santo André - SP, 09280-560}

\author{Maron F. Anka}
\email{maron.anka@fbest.org.br}
\affiliation{QuIIN - Quantum Industrial Innovation, EMBRAPII CIMATEC Competence Center in Quantum Technologies, SENAI CIMATEC, Av. Orlando Gomes 1845, Salvador, BA, Brazil, CEP 41650-010.}

\author{Clebson Cruz}
\email{clebson.cruz@ufob.edu.br}
\affiliation{Centro de Ciências Exatas e das Tecnologias, Universidade Federal do Oeste da Bahia - Campus Reitor Edgard Santos. Rua Bertioga, 892, Morada Nobre I, 47810-059 Barreiras, Bahia, Brasil.}

%%%%%% Abstract %%%%%%
\begin{abstract}
Understanding the thermodynamic properties of quantum systems is essential for developing energy-efficient quantum technologies. In this regard, this work explores the application of quantum computational methods to study the quantum properties and work extraction processes in a dinuclear quantum battery model. 
Our results demonstrate that variational quantum algorithms can reproduce key trends in experimental data, making it possible to analyze the effectiveness of the presented protocol in noisy environments and providing insights into the feasibility of quantum batteries in near-term devices. We have shown that the presence of a noisy environment hinders the accuracy of the evaluation of the amount of energy stored in the system.  Additionally, we analyze the work extraction precision, revealing that although the system can store energy at room temperature, the protocol is highly precise only at low temperatures, and its accuracy at ambient conditions remains limited, compromising its usability.
%Using variational quantum algorithms, we simulated thermal states and implemented an optimal work extraction protocol. 
%and ergotropy under different simulation conditions and highlighting the impact of noise on the battery performance. 
%This research contributes to the broader understanding of quantum thermal devices underscoring the potential of quantum computing in the study of energy storage technologies and materials science.
\end{abstract}
%%%%%% Main Text %%%%%%
\maketitle

\section{Introduction}

The emergence of new technologies based on quantum systems, e.g., quantum computation \cite{gyongyosi2019survey} and quantum cryptography \cite{zhang2024continuous}, is a direct result of advanced research on the quantum properties of composed systems.  This progress marks what is known as the second quantum revolution \cite{mohseni2017commercialize,jaeger2018second,atzori2019second,deutsch2020harnessing}, harnessing quantum phenomena such as coherence and quantum correlations to enable advanced technologies. This progress highlights the importance of understanding how quantum properties, e.g., coherence and quantum correlations, affect energy costs and thermodynamic behavior in quantum devices \cite{giorgi2015correlation,cruz2020quantifying,cruz2022quantum,auffeves2022quantum}. In this scenario, quantum thermodynamics provides an effective framework for such investigations \cite{myers2022quantum}.

Over the past decade, quantum thermodynamics has attracted considerable interest across various scientific communities, including quantum optics, open quantum systems, classical stochastic thermodynamics, quantum information theory, and condensed matter physics \cite{myers2022quantum,sagawa2012thermodynamics,binder2018thermodynamics,deffner2019quantum,gomes2024quantum}. From fundamental theoretical to experimental research, quantum thermodynamics has yielded significant insights, such as the role of information in the entropy production \cite{landi2021irreversible}, fluctuation theorems \cite{seifert2012stochastic,aaberg2018fully}, and thermodynamic uncertainty relations \cite{timpanaro2024family}. Among other topics, quantum thermal devices have been extensively studied in several scenarios, including quantum thermal machines in quasi-static \cite{quan2007quantum,quan2009quantum} and finite-time limits \cite{deng2018superadiabatic,camati2019coherence,das2020quantum} and other modes of operation \cite{de2021efficiency,cherubim2022nonadiabatic,cruz2023quantum}, measurement-based devices \cite{hayashi2017measurement,ding2018measurement,anka2021measurement,malavazi2024weak}, and optimal work extraction devices \cite{PhysRevE.109.044119}, i.e., quantum battery (QB) \cite{campaioli2024colloquium}.

These quantum thermal devices have been proposed using condensed matter physics, especially the quantum battery, which is a device that stores energy in a quantum system to be used for some useful task. For instance, it was proposed using superconducting systems \cite{strambini2020josephson} and spin chains \cite{le2018spin,rossini2019many,zhao2021quantum}. Despite the advances in the theoretical framework, there are challenges to be overcome before we see a practical application of QBs, such as storing energy at room temperature, offering non-destructive access to the stored work, and a charge lifetime comparable with conventional classical batteries.

{Recent studies have demonstrated the experimental viability of QBs using distinct physical platforms, such as microcavities enclosing molecular dyes \cite{doi:10.1126/sciadv.abk3160}, star-topology NMR spin systems \cite{PhysRevA.106.042601}, and multi-layered organic microcavities in scalable room-temperature implementation \cite{hymas2025experimental}.} In parallel, low-dimensional metal complexes (LDMC) emerge as a promising platform for QBs implementations. These advanced materials exhibit unique properties, making them highly suitable for experimental quantum devices. For instance, their molecular structure shields them against environmental perturbations, e.g., high-temperature \cite{reis2012evidence,vcenvcarikova2020unconventional}, magnetic fields \cite{vcenvcarikova2020unconventional,cruz2020quantifying,souza2008experimental}, and pressure \cite{cruz2020quantifying,cruz2017influence}. Another useful feature is that LDMC presents an effective behavior of a two-qubit system that can hold stable quantum correlations above room temperature \cite{cruz2016carboxylate,souza2009entanglement,reis2012evidence}. In this context, some of us proposed quantum thermal devices that implement these advanced materials for quantum batteries at room temperature, showing that it is possible to extract work from genuine quantum correlations \cite{cruz2022quantum}, and a mechanical model for the quantum Stirling cycle \cite{cruz2023quantum,gomes2024quantum}. {The core models used in these works \cite{cruz2022quantum,cruz2023quantum,cruz2016carboxylate,souza2009entanglement,gomes2024quantum,reis2012evidence} are based on dinuclear metal complexes, i.e., a simple magnetic structure composed of two coupled metallic centers, composed by magnetic ions linked directly or through intermediate chemical groups (ligands) \cite{mario_book}.}

{In this scenario, theoretical investigations become increasingly crucial to guide and complement these experimental progresses \cite{razzoli2025cyclic}. 
For this purpose, quantum simulations offer a powerful tool able to explore the dynamical behavior of these systems under realistic conditions, allowing one to assess the impact of decoherence, dissipation, and finite-temperature effects on device performance \cite{RevModPhys.86.153, lloyd1996universal}. In this context, Variational Quantum Algorithms (VQAs) \cite{cerezo2021variational,mcclean2016theory} have emerged as a promising approach to address this complexity, leveraging the potential of quantum computers to simulate quantum systems more efficiently \cite{kandala2017hardware}, especially considering the limitations of NISQ-era devices \cite{preskill2018quantum}. Moreover, the potentialities of quantum variational schemes were explored in ref. \cite{razzoli2025cyclic}, where the authors simulated a cyclic solid-state quantum battery on superconducting IBM quantum machines.}

In this work, we explore a quantum battery model based on quantum discord using Variational Quantum Algorithms simulated using IBM Qiskit \cite{qiskit_website}. We simulate the optimal work extraction protocol proposed in \cite{cruz2022quantum}{, for a two-cell (dinuclear) quantum battery,} in both noiseless and noisy conditions, and compare the numerical results with theoretical and experimental predictions. Our findings illustrated that, despite deviations induced by the noisy environments presented, the variational approach accurately captures the expected relationship between ergotropy and quantum correlations. Unlike the experimental method in \cite{cruz2022quantum}, which infers stored energy via magnetic susceptibility measurements, our approach provides a computational framework for simulating and giving insights into optimizing work extraction in a dinuclear quantum battery model. Additionally, we expand the result presented in reference \cite{cruz2022quantum}, analyzing how the noisy environment influences work fluctuations in the work extraction process, providing a perspective on the practical challenges of implementing quantum batteries with near-term quantum hardware. The results show that, although reference \cite{cruz2022quantum} demonstrated that the system is capable of energy storage at ambient conditions, our simulation of the work extraction protocol reveals satisfactory precision at lower temperatures, while the extraction accuracy at higher temperatures remains a limiting factor for its practical usability.

The paper is organized as follows. Section \ref{sec:2} introduces the theoretical model of the dinuclear quantum battery, highlighting its key properties and the physical principles that govern the optimal extraction of work. Section \ref{compframe} describes the variational quantum algorithm used to simulate thermal states and evaluate work extraction, emphasizing its suitability for noisy intermediate-scale quantum (NISQ) devices. This section also presents the simulation results, including comparisons with theoretical predictions and the effects of noise on performance. Finally, Section \ref{sec:conclusion} provides concluding remarks, summarizing the key findings and their practical implications, and highlighting future research directions.

\section{Theoretical foundations}\label{sec:2}

In order to analyze the work extraction in a dinuclear quantum battery, it is important to establish the theoretical framework governing the working substance. In this context, this section aims to review the fundamental physical model of the system. First, the Hamiltonian framework of the dinuclear quantum battery will be presented, emphasizing the role of spin interactions and the contribution of the external field to its energy eigenstates. The discussion then advances to the concept of ergotropy, quantifying the amount of extractable work from the quantum system of interest. Finally, the optimal work extraction protocol is reviewed, expanding the consolidated references by adding an analysis of the associated energy fluctuations, providing the necessary groundwork for the subsequent quantum simulations. Thus, this section serves as the background for the variational quantum algorithm simulations, analyzing the performance of the system as a quantum battery under realistic conditions of a noisy quantum processor.

\subsection{Working Substance} 

Reference \cite{cruz2022quantum}  developed a theoretical model for a two-cell quantum battery based on a metal-organic copper (II) compound - Cu$_2$(HCOO)$_4$(HCOOH)$_2$(C$_4$H$_{10}$N$_2$). The reduced magnetic unit is formed by two metallic centers of Cu(II), with electronic configuration d$^9$ and $s = 1/2$, with a short intermolecular separation, characteristic of a carboxylate-based metal complex \cite{cruz2016carboxylate}. Such property provides a nearly ideal realization of an isolated two-qubit system \cite{yurishchev2011quantum,gaita2019molecular,moreno2018molecular}, since their intramolecular interaction energy is extremely large when compared to their intermolecular ones \cite{cruz2016carboxylate,yurishchev2011quantum,gaita2019molecular,moreno2018molecular}, which shields the quantum properties of the system against high-temperature effects, enabling the system to store energy throughout quantum correlations at room temperature.

The Hamiltonian model of this system can be described by:
\begin{equation}\label{hamiltonian}
\mathcal{H} = E_0 (S_1^{z} + S_2^{z}) + J (\vec{S}_1 \cdot \vec{S}_2) ,
\end{equation}
where $S_n^{k} = (1/2) \sigma_n^k$, with $\sigma_n^k$ is the Pauli matrix of the {$n-th$} spin and $k \in \{x, y, z\}$. The first term is the Zeeman Hamiltonian {$\mathcal{H}_0= E_0 (S_1^{z} + S_2^{z})$}, where $E_0 = \mu_B g_z B_z$, $g_z$ is the isotropic Landé factor, $\mu_B$ is the Bohr magneton, and $B_z$ is the external magnetic field, describing the energy levels of each cell of the battery: $\mathcal{H}_0 \ket{\epsilon_i} = \epsilon_i \ket{\epsilon_i}$, with associated eigenvalues $\epsilon_1 = -E_0$, $\epsilon_2 = \epsilon_3 = 0$, and $\epsilon_4 = E_0$, and respective eigenstates $\ket{\epsilon_1} = \ket{0 0}$, $\ket{\epsilon_2} = \ket{1 0}$, $\ket{\epsilon_3} = \ket{0 1}$, and $\ket{\epsilon_4} = \ket{1 1}$ defined by the local basis \cite{cruz2022quantum}. The second term is the Heisenberg Hamiltonian, which accounts for the internal interaction between Cu(II) ions, while $J$ represents the magnetic coupling constant. The energy structure of this Hamiltonian is composed of a singlet state $\ket{\beta_-}$ with energy $E_- = -3J/4$ and a triplet-degenerate subspace $\{\ket{\beta_t} \} = \{\ket{\beta_+}, \ket{0 0}, \ket{1 1}\}$, with energy $E_t = J/4$ and {$\ket{\beta_{\pm}} = (\ket{0 1} \pm \ket{1 0})/\sqrt{2}$}. %It is important to note that the energetic difference between the ground and first excited states, $\Delta E = E_t - E_- =J > 0$, leads to an entangled singlet ground state with anti-parallel alignment.

The magnetic coupling for this specific material is huge, because of the very short intermolecular separation, which was experimentally verified to be $J/k_B = 748$ K \cite{cruz2016carboxylate,cruz2022quantum}. This value enables the existence of stable quantum correlations above room temperature in this material, experimentally reported in reference \cite{cruz2016carboxylate}. Another consequence of such a large coupling value is that the quantum level crossing induced by the magnetic field $B_z$, which leads to population inversion, occurs only for non-experimental values of a critical field $B_c \approx 556 T$. Therefore, this compound behaves as an effective two-level system with ground and excited states given by $\ket{\beta_-}$ and $\beta_t$, respectively, for $E_0 \ll J$.

\subsection{Ergotropy}

The concept of ergotropy was first proposed in Ref.\cite{allahverdyan2004maximal} as the maximum amount of energy extractable from a quantum system via cyclic unitary transformations \cite{alicki2013entanglement,campaioli2024colloquium}. Much effort has been put into the investigation of ergotropy in different scenarios, for example, entanglement \cite{alicki2013entanglement,kamin2020entanglement,liu2021entanglement} and quantum coherence \cite{ccakmak2020ergotropy,francica2020quantum,sone2021quantum,francica2022quantum,niu2024experimental} as resources, open systems framework \cite{gherardini2020stabilizing,carrega2020dissipative,xu2021enhancing}, and charging/discharging processes \cite{friis2018precision,kamin2020non,ghosh2021fast,hu2022optimal}.
\begin{comment}
The extraction of work from a quantum system is a cyclic unitary operation generated by a time-dependent Hamiltonian $H_t = H + H_{ext}(t)$, where $H$ is the system's Hamiltonian and $H_{ext}(t)$ is the time-dependent coupling with an external agent where the work is deposited. The necessary condition for a cyclic process, with duration $\tau$, is given by $H_{ext}(0) = H_{ext}(\tau) = 0$. After unitary evolution, the amount of energy extracted in the form of work \cite{allahverdyan2004maximal}, is given by

\begin{equation}
\mathcal{W}(T,B_z) = \operatorname{tr}[\rho(T,B_z) H_0] - \operatorname{tr}[V \rho(T,B_z) V^{\dagger} H_0],
\end{equation}
where $\rho(T,B_z)$ is the initial state of the system, $H_0$ is the Hamiltonian reference of the battery, and $V$ is the unitary operator responsible for the work extraction.
\end{comment}

The ergotropy can be written as the maximum work that can be extracted from a quantum system via unitary operation {$U$}:
{
\begin{equation}
\begin{split}
\mathcal{E}(T,B_z) &= %\mbox{max}_{U \in \mathcal{U}} \{ \mathcal{W}(T,B_z)\} \\
\operatorname{tr}[\rho(T,B_z) \mathcal{H}_0] - \mbox{min}_{U \in \mathcal{U}} \{\operatorname{tr}[U \rho(T,B_z) U^{\dagger} \mathcal{H}_0]\},
\end{split}
\label{ergotropy}
\end{equation}}
{where the minimization} is taken over the set {$\mathcal{U}$} of all unitary operators. From this equation, it is clear that not all internal energy ($E$) can be extracted as work, since $\mathcal{E}(T,B_z) = E(\rho(T,B_z)) - E_0$, where $E(\rho(T,B_z))$ is the internal energy of the initial system \cite{allahverdyan2004maximal,niedenzu2019concepts}.
{It was shown that, by ordering the eigenvalues of the self-Hamiltonian in increasing order, $\epsilon_1 \leq \epsilon_2 \leq \cdots \leq \epsilon_N$, and the eigenvalues of the density matrix of the system in decreasing order, $\varrho_1 \geq \varrho_1 \geq \cdots \geq \varrho_1$, we can write the ergotropy as \cite{allahverdyan2004maximal}}

{\begin{equation} \label{ergotropy}
\mathcal{E}(T,B_z) = \sum_{i,j}^{N,N} \varrho_i \epsilon_j ( |\langle\varrho_i \vert \epsilon_j\rangle |^2 - \delta_{i,j}).
\end{equation}}

It is worth noting that we are assuming that the initial state of the system is a thermal state given by $\rho(T, B_z) = e^{-\mathcal{H}/k_B T}/tr[e^{-\mathcal{H}/k_B T}]$, which is diagonal in the coupled basis $\{\ket{\varrho_i}\}$. Thus, it is a complete passive state, from which no energy can be extracted. However, we consider the density matrix in the uncoupled basis of the self-Hamiltonian of the battery $\{\ket{\epsilon_j}\}$, from which it takes the form of an active X-state, in which a finite amount of energy can be extracted:
\begin{equation}
               \rho(T,B_z) = \dfrac{e^{\xi}}{2 Z} \left[
\begin{matrix}
2 e^{\beta E_0} & 0 & 0 & 0  \\
0 & 1 + e^{-4 \xi} & 1 - e^{-4 \xi} & 0 \\
0 & 1 - e^{-4 \xi} & 1 + e^{-4 \xi} & 0 \\
0 & 0 & 0 & 2 e^{-\beta E_0}
\end{matrix} \right],
\end{equation}
where $Z = e^{\xi} + e^{-3 \xi} + 2 e^{\xi} \cosh{(\beta E_0)}$ is the partition function, $\beta = 1/k_B T$, and $\xi = -\beta J/4$. %Thus, it is an active state concerning the Hamiltonian of the battery.

\subsection{Optimal Work Extraction}

In this scenario, Ref.\cite{cruz2022quantum} introduces the evaluation of the ergotropy through an experimental measure of a macroscopic property, the magnetic susceptibility $\chi(T)$. This is achievable for the regime of magnetic susceptibility, $E_0 << k_B T$. {In this limit, using Eq.(\ref{ergotropy}),  we can write the ergotropy as a function of the temperature as \cite{cruz2022quantum}}
\begin{equation}\label{ergosus}
\mathcal{E}_{E_0 << k_B T} (T) = E_0 \dfrac{k_B T\chi(T)}{4 N_A g^2 \mu_B^2} (e^{-4 \xi} - 1),
\end{equation}
in terms of $\chi(T)$ given by the Bleaney-Bowers equation ~\cite{bleaney1952anomalous}
		\begin{align}
		\chi(T) &=\frac{2N(g\mu_B)^2}{k_B T}\left(\frac{1}{3+e^{-{4}\xi}}\right).
		\label{bleaney}
		\end{align}
where $N_A$ is the Avogadro number. In this regard, Eq.  \eqref{ergosus} can be written directly proportional to the quantum discord of the system \cite{cruz2022quantum}:
\begin{equation} \label{eq:erg}
\mathcal{E}_{E_0 << k_B T} (T) = 2 E_0 \mathcal{D}, \forall J > 0,
\end{equation}
where 
\begin{equation} \label{eq:disc}
\mathcal{D} = \dfrac{1}{2} \Bigg| \dfrac{2 k_B T}{N_A g^2 \mu_B^2} \chi(T) - 1 \Bigg|
\end{equation}
is the Schatten one-norm quantum discord \cite{cruz2016carboxylate} \footnote{{More details on the general form of the ergotropy and quantum discord, and the derivation of Eqs.(\ref{ergosus}-\ref{eq:disc}) can be found in the appendix of Ref.\cite{cruz2022quantum}.}}.

The work extraction protocol is characterized by the unitary operator $U = \sum_i \ket{\varepsilon_i}\bra{\varrho_i}$ \cite{allahverdyan2004maximal}, such that the initial active state becomes a passive state, $\rho(T,B_z) \rightarrow \sigma(T,B_z) = U \rho(T,B_z) U^{\dagger}$, where $[\sigma(T,B_z),\mathcal{H}_0] =0$. In the studied case, in the low-temperature limit $k_B T \ll J$, the maximum ergotropy state is given by $\rho = \ket{\beta_-}\bra{\beta_-}$, while the empty state, with zero ergotropy, is given by $\sigma = \ket{0 0}\bra{0 0}$. This optimal work extraction is experimentally achievable in the thermal range of $T < 83$ K \cite{cruz2022quantum}. 
{At this temperature, the system reaches the pure state regime. Thus, the $83$ K temperature represents a thermodynamic threshold which marks the transition from a regime of maximized quantum resources \cite{cruz2016carboxylate}. Above this temperature, the thermal energy becomes significant to populate excited states, causing the loss of the purely ground-state occupancy. The emergence of this threshold temperature is attributed to the magnetic coupling characteristics inherent to dinuclear metal complexes, particularly those based on carboxylate groups \cite{cruz2016carboxylate,cruz2022quantum}.  In such materials, a strong metal-to-metal magnetic interaction, mediated through carboxylate bridges with a syn–syn conformation  \cite{cruz2016carboxylate}, leads to a significant energy gap between the singlet ground state and the first excited triplet states \cite{cruz2022quantum}. Therefore, below this temperature, the system predominantly occupies the maximally entangled singlet ground state, maximizing quantum correlations and leading to an optimal work extraction in a quantum battery model \cite{cruz2022quantum}.}

Therefore, in the quantum computing framework, the unitary operator responsible for this operation can be implemented by a sequence of logical gates as follows:
\begin{equation}
U \ket{\beta_-} = (H_1 \otimes 1_2) CNOT_{12} \big( \dfrac{\ket{1 0} - \ket{0 1}}{\sqrt{2}} \big) = \ket{0 0}.
\end{equation}

{This sequence first applies a CNOT gate with qubit 1 as control and qubit 2 as target ($CNOT_{12}$), followed by a Hadamard gate on qubit 1 ($H_1$).} Thus, this optimal work extraction can be tested by exploring a quantum computing simulation as a proof of principle for the proposed quantum battery model using a quantum computer to implement this work extraction protocol.

\subsection{Protocol Precision}
 
Before presenting the quantum computing framework for the energy extraction protocol, we briefly expand the result presented in reference \cite{cruz2022quantum} by discussing a figure of merit for analyzing the charging/discharging process in quantum batteries. In Ref.\cite{friis2018precision}, the authors proposed a way to evaluate the precision in the charging process by analyzing the average energy increase after the unitary operation. To charge an initial thermal state $\tau$ by unitary operation, such that $\rho = U \tau U^{\dagger}$, an amount of average energy must be increased, leading to $\Delta E = E(\rho) - E(\tau)$, where $E(\rho) = \operatorname{tr}[\rho \mathcal{H}_0]$. The precision of this process is evaluated by an increase in the standard deviation of the system Hamiltonian:

\begin{equation}
\Delta \sigma = \sqrt{V(\rho)} - \sqrt{V(\tau)},
\label{variance}
\end{equation}
where $V(\rho) = (\Delta \mathcal{H}_0(\rho))^2 = \operatorname{tr}(\mathcal{H}_0^2 \rho) - (\operatorname{tr}(\mathcal{H}_0 \rho))^2$. The smaller variance corresponds to a higher precision of the charging/discharging process. Thus, Eq. (\ref{variance}) serves as a parameter for estimating the confidence in the implemented energy extraction protocol \cite{friis2018precision}.

In our case, we adopt the same theoretical approach to analyze the discharging process. We are interested in applying a unitary operation in which the average energy is decreased by $\Delta E$, extracting energy from the active state $\rho(T, B_z)$ leading to the passive state $\sigma(T, B_z)$. While in the charging process the difference between the mean values of energy is not the ergotropy, since no optimization in the unitary operator is necessary, here we get $\mathcal{E}(T,B_z) = \Delta E = E(\rho(T, B_z) - E(\sigma(T, B_z)))$. Applying this to the optimal protocol described previously, we evaluate the precision of the extracted work protocol by evaluating Eq.(\ref{variance}). 

\section{Energy extraction protocol using a variational quantum algorithm}
\label{compframe}

Simulating quantum properties in complex systems, such as metal complexes, is a significant challenge in computational chemistry and condensed matter physics \cite{articlea}. These systems exhibit intricate electronic interactions that often demand methods capable of capturing their quantum nature beyond classical computational capabilities \cite{Zobel_2021}. {An alternative with the potential to avoid this problem
is the Variational Quantum Algorithms (VQAs): a hybrid computational framework that combines the strengths of classical and quantum devices.}

{Variational Quantum Algorithms (VQAs) represent one of the most promising approaches to harness the potential of intermediate-scale quantum devices, known as NISQ (Noisy Intermediate-Scale Quantum) devices \cite{Bharti_2022}. In this context, VQAs have been widely applied to solve a variety of relevant problems, such as computing energy spectra in molecular systems \cite{kandala2017hardware}, combinatorial optimization \cite{farhi2014}, and training machine learning models \cite{Mitarai_2018}. Among these applications, quantum system simulation stands out as a field of great interest, encompassing everything from modeling interactions in strongly correlated systems \cite{McArdle2020,kandala2017hardware} to exploring dynamics in out-of-equilibrium systems \cite{Yuan_2019}. }

{In a typical VQA, the main idea is to encode the quantum properties of the problem into a parameterized quantum circuit (ansatz) \cite{cerezo2021variational}, which represents the state of the system under study. This ansatz is iteratively optimized by minimizing a cost function that encodes the system's physical properties, such as its energy. Classical optimization algorithms guide this iterative process, adjusting the parameters to find the extrema of the cost function, which correspond to the solution of the problem.}

{A well-known example of this approach is the simulation of the ground state energy of a simple molecule like $H_2$. In this case, a quantum circuit prepares a trial state, the system's energy is measured, and a classical optimizer adjusts the circuit parameters to minimize that energy. Repetition of this process converges the algorithm toward the ground state of the molecule, demonstrating the effectiveness of VQAs in capturing quantum behavior even on noisy quantum devices with few qubits. This simple case has already been demonstrated experimentally \cite{Peruzzo_2014}. }

{In a previous work, we used a VQA, the Variational Quantum Thermalizer algorithm (VQT),  to investigate the thermal properties of dinuclear metal complexes based on the Hamiltonian of the system \cite{dasNevesSilva_2024}.} This involves the preparation of thermal states that encode information about the properties of the complex at finite temperatures. Simulating these thermal states makes it possible to extract quantities such as specific heat and magnetic susceptibility.

{Several alternative methods for preparing thermal states exist, including quantum Metropolis sampling \cite{temme2011quantum}, Grover-based algorithms \cite{chiang2010quantum}, and Thermal Pure Quantum (TPQ) states \cite{sugiura2013canonical}. However, they often require deep circuits or large systems, making them unsuitable for NISQ devices. Variational methods are more compatible with near-term hardware. Among them, VQT stands out by optimizing a physically motivated cost function, the quantum free energy, which naturally balances energy minimization and entropy maximization. This avoids simulating non-unitary dynamics (as in variational imaginary-time evolution \cite{mcardle2019variational}) and enables direct convergence to thermal equilibrium. 
}

In this section, we present the results of simulating the quantum properties of a dinuclear metal complex using a variational quantum algorithm (VQA). We first outline the algorithmic framework and computational methodology employed to model the system, emphasizing the suitability of VQAs for such tasks. Next, we detail the simulation outcomes, focusing on key quantum metrics such as discord and ergotropy. Additionally, we apply the optimal work extraction protocol to evaluate the properties associated with the process. These findings underscore the capability of VQAs to elucidate the quantum behavior of complex systems, offering a bridge between theoretical predictions and practical applications.

\subsection{Variational quantum thermalizer and protocol of optimal work extraction }

In this work, we employ the Variational Quantum Thermalizer (VQT) algorithm \cite{verdon2019quantum,dasNevesSilva_2024} to approximate the thermal states of the Hamiltonian described in Eq. (\ref{hamiltonian}) \cite{dasNevesSilva_2024}. %As detailed in Section \ref{sec:2}, due to the dominance of the magnetic coupling constant $J$ of the metal complex ($ J \gg E_0 $), we focus solely on the leading term of the Hamiltonian, $ J (\vec{S}_1 \cdot \vec{S}_2)$, which represents the magnetic exchange interaction between the spins, with $ J/k_B = 748~\text{K} $.
To approximate the thermal states, we utilize a probabilistic quantum ansatz \cite{verdon2019quantum} $\rho_{\theta, \phi}$, parameterized by variables $\theta$ and $\phi$.  {These parameters define both the initial state preparation and the structure of the quantum circuit used in the algorithm. Specifically, $\theta$ governs the classical probability distribution $p_\theta$ from which computational basis states are sampled, and $\phi$ parameterizes the unitary operations $U(\phi)$ that act on these states via a variational quantum circuit.}

%These parameters define the structure of the quantum circuit and the corresponding probabilistic distribution, allowing the ansatz to represent mixed quantum states. Specifically, $\phi$ parameterizes the unitary transformations applied to the quantum circuit. At the same time, $\theta$ governs the distribution of probabilities responsible for preparing the initial states $\rho^{(i)}_{\theta}$, before being processed by the circuit parameterized by $\phi$.

\begin{figure*} 
    \centering
    \includegraphics[width=0.8\linewidth]{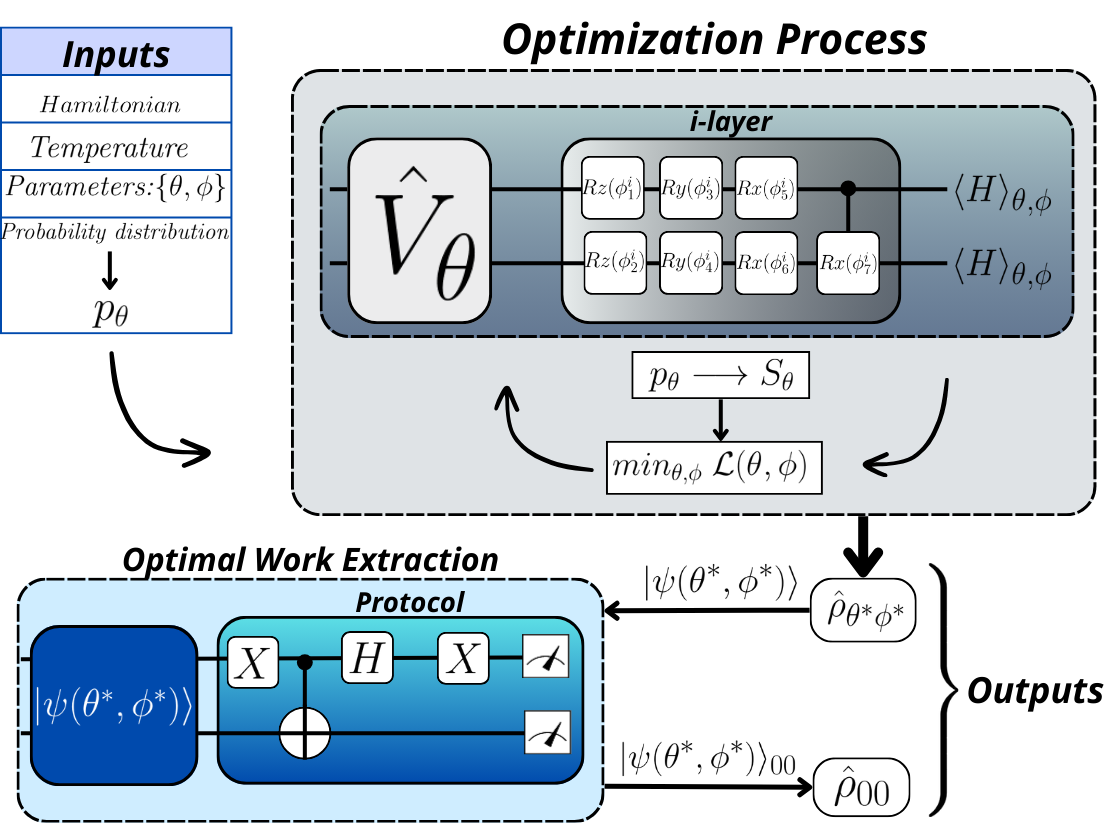}
    \caption{ \justifying  Algorithm Workflow. The top circuit refers to the dynamics of the Variational Quantum Thermalizer to simulate the state of interest, resulting in a set of density operators $\rho^{(i)}$ related to different temperatures. Then, the optimized state $\ket{\psi (\theta^*, \phi^*)}$ is forwarded to the bottom circuit below, where the optimal work extraction protocol is applied, taking the system to a passive state $\ket{\psi (\theta^*, \phi^*)}_{0 0}$. }
    \label{fig:VQT}
\end{figure*}

{In our implementation, schematized in Fig. \ref{fig:VQT}, the block $\hat{V}_\theta$ prepares the initial mixed state  $\rho^{(i)}_{\theta}$, which encodes a classical probability distribution determined by the parameters $\theta$. This state serves as the input for the variational quantum circuit. The unitary operator $U(\phi)$ is composed of a sequence of parameterized single-qubit rotation gates, including $R_x(\phi)$, $R_y(\phi)$ and $R_z(\phi)$ (rotation around the $x$, $y$ and $z$-axis), along with two-qubit entangling gates such as the controlled-NOT (CNOT). These gates are arranged in alternating layers across the qubits to form the unitary transformation $U(\phi)$, which maps the initial state into a variationally optimized quantum state. The parameter set $\phi$ determines the rotation angles and defines the structure of each variational layer.}

{The goal of the algorithm is to find optimal parameters $\theta$ and $\phi$ that minimize the quantum relative free energy \cite{verdon2019quantum} of the ansatz concerning the target thermal state:}:
\begin{equation} \label{eq:cost}
    \mathcal{L}_{\theta,\phi} =   \beta \text{tr}(\rho_{\theta,\phi}\mathcal{H}) - S(\rho_{\theta,\phi})~,   
\end{equation}
where $\text{tr}(\rho_{\theta, \phi} H)$ represents the expected energy of the ansatz, and $S(\rho_{\theta,\phi})$ denotes its Von Neumann entropy. This cost function balances energy minimization and entropy maximization, guiding optimization toward thermal equilibrium. By minimizing $\mathcal{L}_{\theta, \phi}$, the algorithm reconstructs the thermal state, enabling extraction of thermodynamic properties such as energy, entropy, and magnetic susceptibility. More details on obtaining these properties,{as well as implementation details,} for dinuclear metal complexes using {VQT} can be found in Ref.\cite{dasNevesSilva_2024}.

{As shown in our previous study, this architecture captures thermodynamic properties of molecular magnetic systems. The clear separation between the latent-state \cite{verdon2019quantum} preparation governed by $\theta$ and the expressive variational circuit defined by $\phi$ plays a crucial role in both efficiency and physical interpretability of the algorithm.}

In this study, we calculate the magnetic susceptibility and investigate its relationship with the ergotropy derived from discord, following the relation presented in Eq. (\ref{ergosus}).
%In this study, we calculate the magnetic susceptibility and investigate its relationship with quantum discord, as described in Eq. (\ref{eq:disc}). Additionally, we explore the ergotropy derived from discord, following the relation presented in Eq. (\ref{eq:erg}).
%After the optimization process, the algorithm output a list of density matrices $\rho^{(i)}$, constructed from the optimized state $\ket{\psi (\theta^*, \phi^*)}^{(i)}$, directly related to specific temperature values $T^{(i)}$. This state is then send to the protocol of optimal work extraction.
Finally, with the list of density matrices $\rho^{(i)}$, obtained after the optimization process and constructed from the optimized state $\ket{\psi (\theta^*, \phi^*)} ^{( i)}$, each state corresponding to a specific temperature value $T^{(i)}$ is subjected to the optimal work extraction protocol.
As discussed, the protocol is built from a sequence of CNOT and Hadamard gates. The NOT gate in the circuit is due to the fact that we consider $\ket{0}$ as the ground state in the simulation, so the CNOT gate was controlled with $\ket{0}$ as the control qubit. As a result, the algorithm output after the protocol is the state of minimal energy. The whole workflow of the algorithm can be visualized in Figure ~\ref{fig:VQT}.

\subsection{Ergotropy and Discord Simulation}

The Ergotropy described in equations \eqref{ergosus} depends on the magnetic susceptibility, which is directly calculated from the density operator $\rho$, obtained from the quantum algorithm (Figure \ref{fig:VQT}). %In order to simulate these proprieties, the algorithm was performed using $110$ different values of temperature in a range of $\{0 \ K - 300 \ K\}$. This choice is due to the fact that we obtained $100$ temperature values from the previous experimental data of our group, which was available in a range of $\{40 \ K - 300 \ K\}$. For the algorithm, we added $10$ extra values from $\{0 \ K - 40 \ K\}$ in order to analyze the behavior of the curve for low temperatures. 
In this context, the algorithm was simulated in a noiseless and a noisy simulator, aiming to analyze its behavior in an ideal and a more realistic environment, respectively. For this purpose, it was used the open-source software development kit from IBM Quantum Experience, or simply qiskit \cite{qiskit_website}. The noisy simulator was modeled considering the fake backends available, with the noise model described in Appendix \ref{app:noise}.

%%%%%%%%%% Mudança de Ordem %%%%%%%%%%%%%
The choice of the ideal algorithm configuration for the optimization process was preceded by a series of tests, considering the performance of the algorithm in simulating the ergotropy obtained from the experimental data \cite{cruz2022quantum}. Thus, we evaluated the effectiveness of using different optimizers and layers based on two criteria:
\begin{itemize}
    \item[(i)] \label{item:avError} The \textit{average error accumulation} in experimental $(\varepsilon_{exp}^{(i)})$ and simulated $(\varepsilon_{sim}^{(i)})$ ergotropy,  given by  $\frac{1}{N}\sum_{i=0}^{N-1}\vert \varepsilon_{exp}^{(i)} - \varepsilon_{sim}^{(i)} \vert $, where $N$ is the number of points, in order to evaluate the fidelity between the curves of ergotropy.
    
    \item[(ii)] \label{item:avEval} The \textit{average function evaluation} of the optimization steps used in the algorithm, considering each point. Let $n^{(i)}$ be the number of evaluations of an optimization process at a point $(i)$, and the average function evaluations is given by $\frac{1}{N}\sum_{i=0}^{N-1}n^{(i)}$.
    
\end{itemize}

\begin{figure*}
 \centering
  \subfloat[]{\includegraphics[width=.5\linewidth]{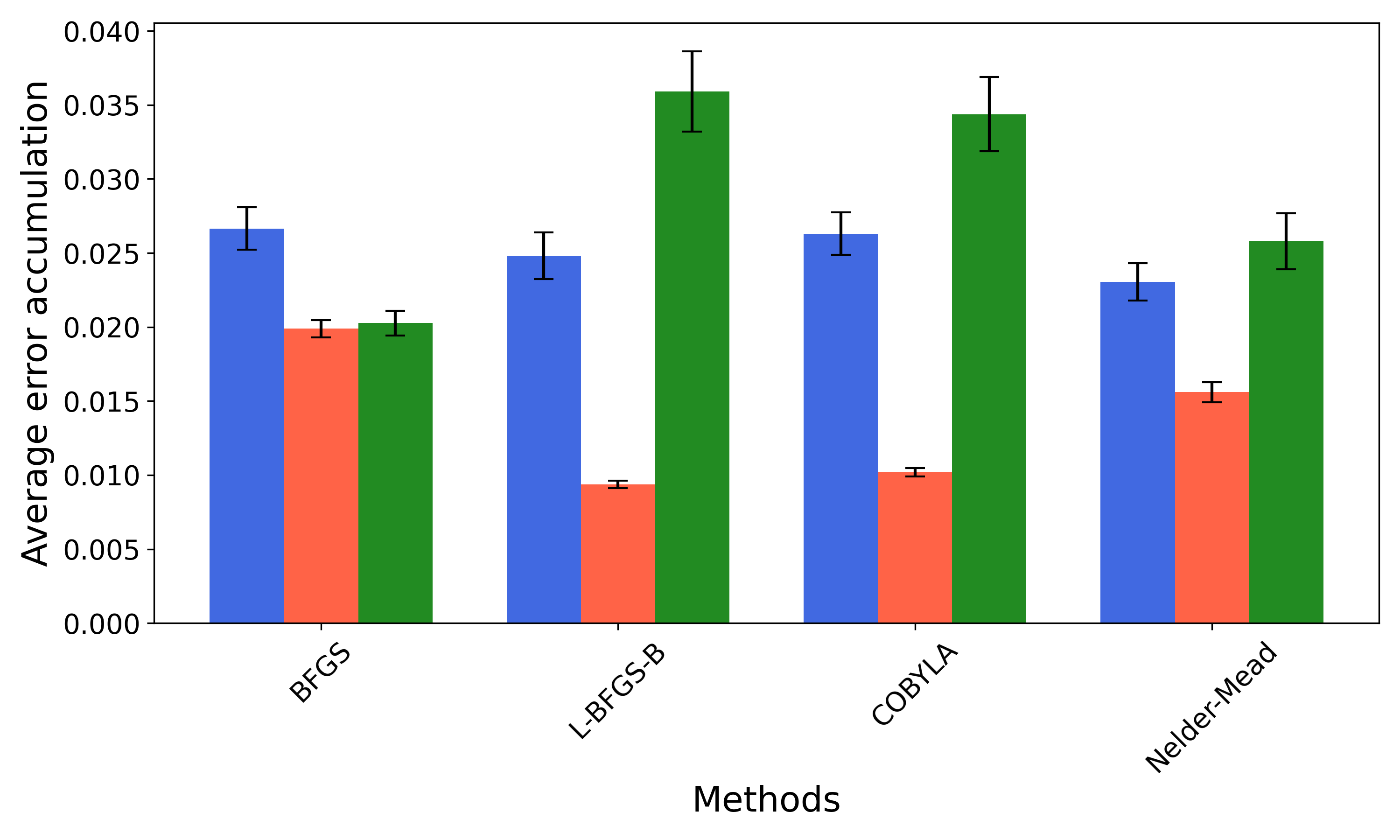}}
  \subfloat[]{\includegraphics[width=.5\linewidth]{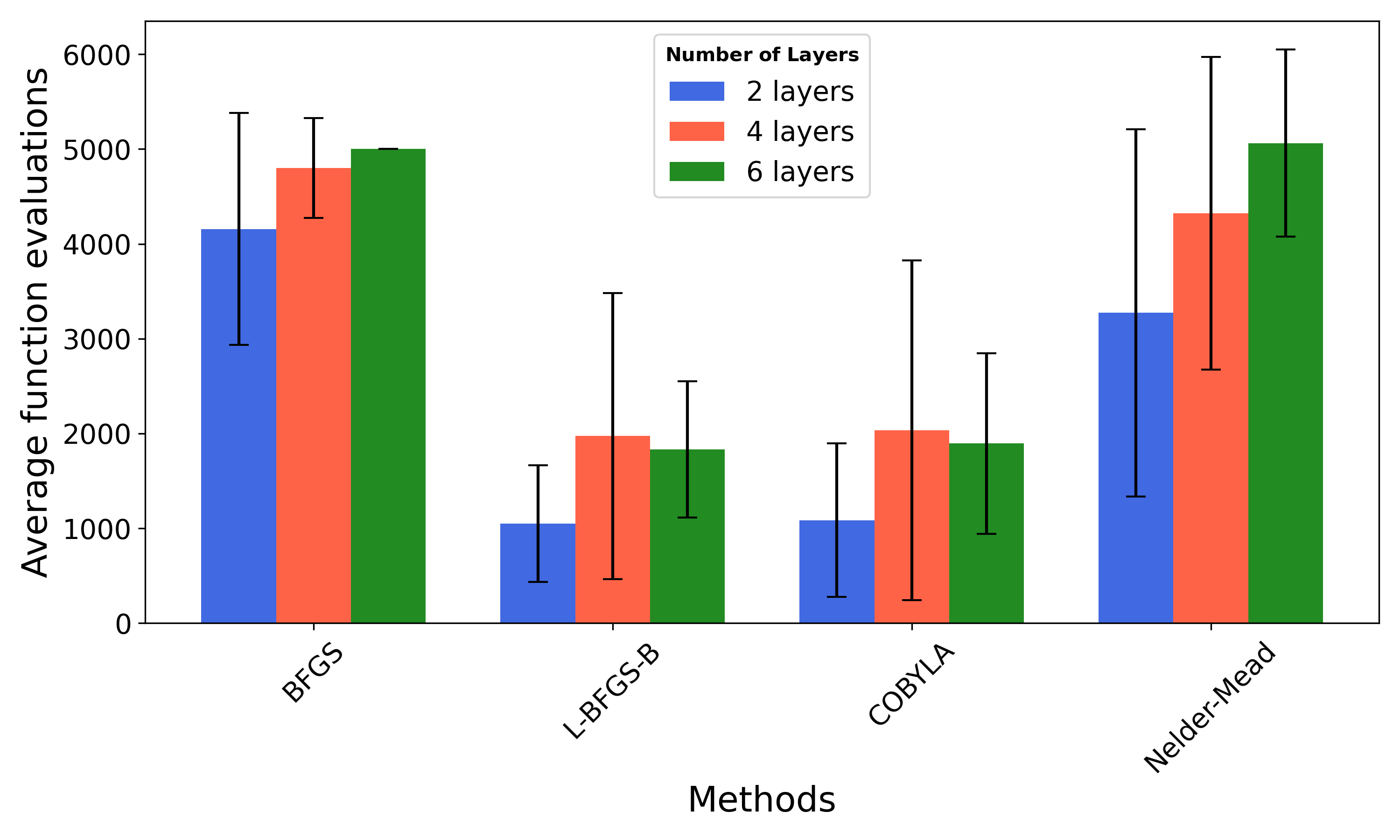}}
  \caption{ \justifying  Simulation test results for different algorithm configurations for four optimizers with 2, 4, and 6 layers considering the average error accumulation defined in {(i) (left-a)} and the average function evaluations define in {(ii) (right-b)}. As can be seen, L-BFGS-B and COBYLA achieved better results in both metrics.}
  \label{fig:comparacao_resultados}
\end{figure*}

While the first one can be seen as a measure to describe the accuracy of the algorithm, the second one can give us partial information about its cost. Considering these metrics, we simulate the algorithm with 2, 4 and 6 layers and four different optimizers:  Constrained Optimization BY Linear Approximation (COBYLA), Nelder-Mead, BFGS and L-BFGS-B \cite{bressert2012scipy}. The COBYLA and Nelder-Mead are straightforward methods, which do not require the calculation of derivatives of the function. They are generally robust, but can converge more slowly than gradient-based methods on softer problems \cite{powell1994advances,cheng2024quantum,singh2023benchmarking,pellow2021comparison}. The BFGS and L-BFGS-B are gradient-based methods that use information about the slope of the function to determine the search direction \cite{Xie2019Analysis, Berahas2017A}. The analysis results can be visualized in Figure \ref{fig:comparacao_resultados}.

In all scenarios, the configuration using COBYLA and L-BFGS-B had better results, which shows that the algorithm can have efficient results with specific gradient-based or free-gradient optimizers. We choose the COBYLA method\footnote{COBYLA is a commonly used derivative‐free numerical optimization algorithm, it was tailored for constrained problems in which gradient information is unavailable. The method involves an interactive strategy continuously refined by the scientific community over the years and has been applied in different hybrid quantum-classical algorithms \cite{cheng2024quantum,singh2023benchmarking,pellow2021comparison}.} optimizer, because it can recover the behavior of the curve for large values of $T$, while the L-BFGS-B diverges. This result agrees with the cost in equation (\ref{eq:cost}), since the distance between the exact state and the ansatz should decrease for large values of $T$.

Note that $4$ layers had better results in all cases, which means that a few layers can not be enough to simulate the whole space domain of the problem, but more layers do not imply more accurate results. Recent works presented that deep parametrized circuits can occasionally find barren plateaus, even using quantum-inspired ansatz, which can explain this behavior \cite{cerezo2023does}. 

%%%%%%%%%%%%%%%%%%%%%%%

\begin{figure}
 \centering
 \includegraphics[width=\columnwidth]{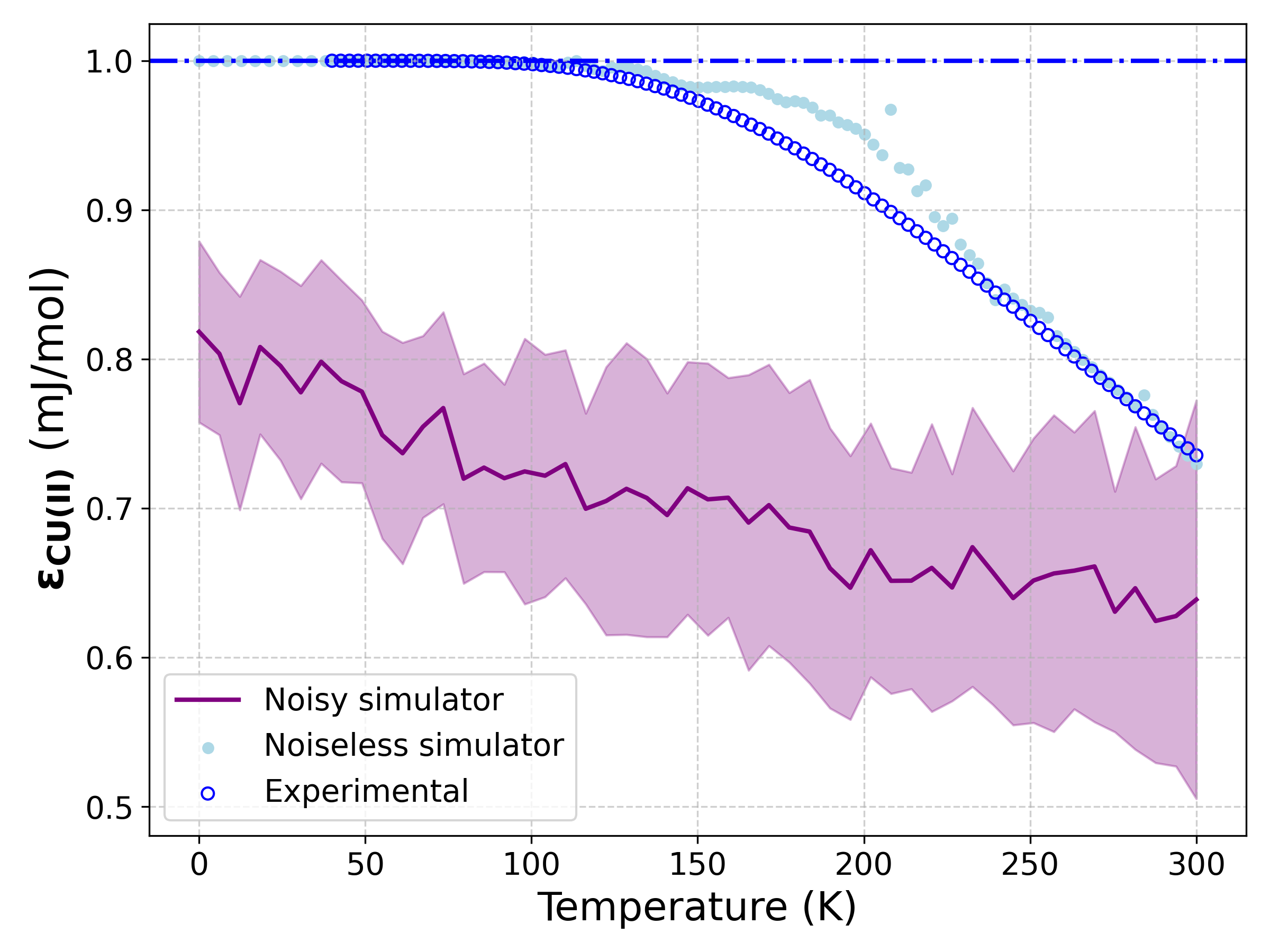}
  %\subfloat{\includegraphics[width=\columnwidth]{ergotropia_vs_temp.png}}
  %\subfloat{\includegraphics[width=\linewidth]{discordia_vs_temp.png}}
  \caption{\justifying Simulation of Ergotropy as a function of the temperature using a noiseless and noisy simulator. Simulated noiseless data (filled points) and a noisy simulation (solid line) were obtained from molar magnetic susceptibility calculations, exploring its relationship with ergotropy as described in Eqs. \eqref{ergosus}. The molar magnetic susceptibility was derived from thermal states approximated by the Variational Quantum Thermalizer (VQT) algorithm, considering temperature values in the range {$0 \ K - 300 \ K$} and using 5000 optimization steps with the COBYLA method. {The noisy simulation was perform 30 times, with a solid line as its average and the shadow as the standard deviation.} {The experimental data was obtained from \cite{cruz2022quantum}, considering the normalized ergotropy}}
  \label{fig:results}
\end{figure}

The results using the start configuration for the noiseless and noisy scenarios can be seen in Figure \ref{fig:results}. Considering this analysis, we perform the algorithm 30 times to extract a considerable average of the data, represented as a solid line, while its standard deviation is represented as a shadow. In both cases, the standard algorithm configuration remained with $4$ layers ansatz and the COBYLA as the optimizer, with a limit of $5000$ optimization steps.

As can be seen, the noiseless simulations show certain consistency with the experimental magnetometry data. The simulation points follow the expected dynamics of ergotropy constant for low temperatures, as expected in theory \cite{cruz2022quantum}. If compared to the experimental data, the algorithm presents a satisfactory accuracy in most of the points, with small oscillations between $\{160 \ K - 230 \ K\}$. This behavior suggests that the algorithm faces intrinsic challenges that lead to discrepancies between simulation results and theoretical or experimental predictions in this particular temperature range.

In this scenario, one of the key obstacles in this variational algorithm approach is the phenomenon of barren plateaus, where the optimization landscape becomes nearly flat, resulting in vanishing gradients that hinder parameter convergence \cite{cerezo2021variational}. This issue is particularly pronounced in deep and complex quantum circuits, where the probability of encountering such flat regions increases, often compounded by noise and hardware limitations \cite{larocca2024review}. This phenomenon can increase error accumulation in the algorithm due to the number of layers. Thus, some recent efforts have sought to circumvent this issue by either using classical shadows \cite{sack2022avoiding}, Bell measurements \cite{cao2024accelerated}, or quantum circuit evolution \cite{franken2022quantum}. 

On the other hand, for the noisy simulation, the results show certain difficulty in converging with the experimental curve throughout the execution, as expected. Unlike the noiseless simulation, where we usually expect optimal results, the noise introduced into the simulator tends to slightly distance the results from the exact value. As a result, the ergotropy does not approach to the experimental curve at any temperature. Thus, the presence of a noisy environment hinders the accuracy of the evaluation of the amount of energy stored in the system. These results demonstrate the limitations of the algorithm for execution on real computers of the NISQ era due to the noise influence \cite{preskill2018quantum, li2023qasmbench}. However, it is important to note that the difference between experimental results and noisy results is relativity small. These values suggest that error mitigation techniques can be used to achieve better results\footnote{A detailed study of error mitigation strategies is beyond the scope of this work and will be addressed in future studies.} \cite{barron2020measurement, botelho2022error, ravi2022vaqem}.

\subsection{Work Extraction Simulation and Protocol Precision}

\begin{figure*}
 \centering
  \subfloat{\includegraphics[width=.5\linewidth]{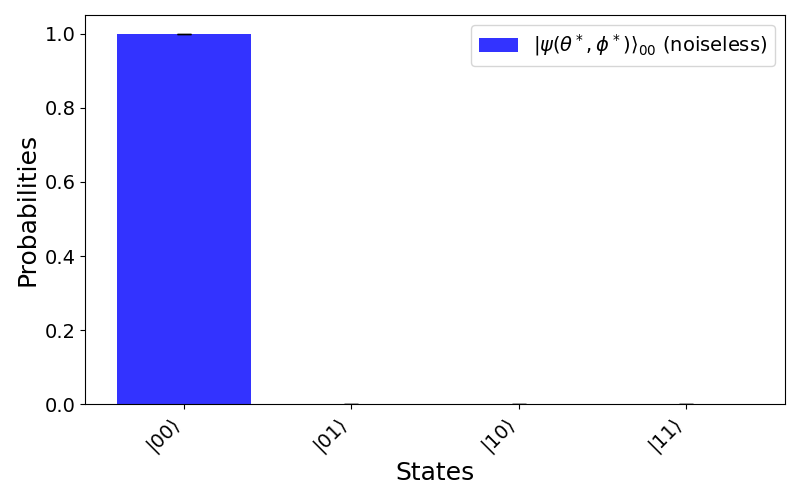}}
  \subfloat{\includegraphics[width=.5\linewidth]{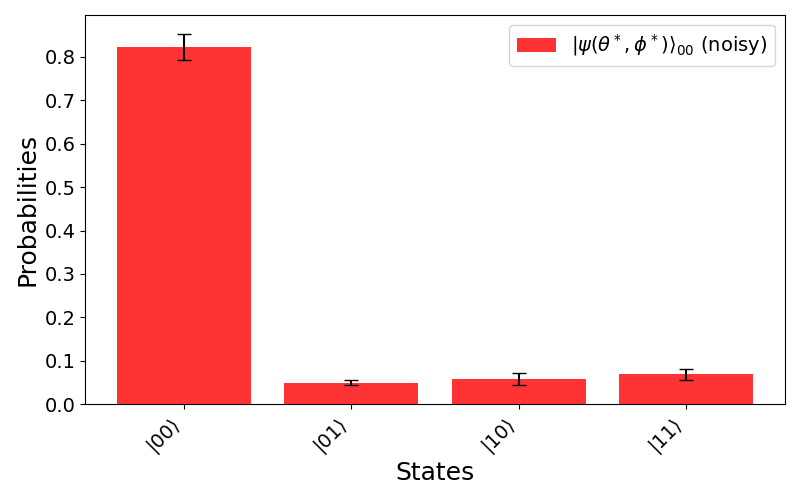}}
  \caption{\justifying  Final state after applying the optimal work extraction protocol for $T < 100 \ K$ in the noiseless simulation (left) and noisy simulator (right). The results show that the protocol allows an efficient extraction of the energy, taking a maximally entangled to a ground state configuration in both cases for low temperatures. The noisy simulation was performed 30 times, with the red bars as its average and the small lines in its top as the standard deviation.}
  \label{fig:pass}
\end{figure*}

In the following, {the protocol for optimal work extraction shown in Fig. \ref{fig:VQT}} is applied to the $\rho$ operator, prepared based on ensemble $\{ p_\theta, \ket{\psi (\theta^*, \phi^*)} \}$, considering the regime of pure states for the system for low temperatures \cite{cruz2022quantum}. This state is sent to the protocol, whereby the system is driven from the maximally entangled ground state $\rho_{\ket{{\beta_{-}}}}$, an active state that stores the maximum amount of energy $E_0$, to the passive state $\rho_{\ket{00}}$, with zero energy stored. This passive state is obtained as a {second output of the algorithm and is constructed from the states $\ket{\psi (\theta^*, \phi^*)}_{0 0}$}. %Figure \ref{fig:results} shows that the maximum amount of stored work can be obtained for temperatures lower than $100 \ K$, so it is expected that the protocol takes $\ket{\psi (\theta^*, \phi^*)} \}$ to a configuration near to the ground state $\ket{00}$. 
Considering this system behavior, we apply the protocol for the states with $T < 100 \ K$ in a noiseless and noisy simulator using the standard configuration, since Fig. \ref{fig:results} shows a maximum ergotropy in this range. The results can be seen as graph bars in Figure \ref{fig:pass}, which shows the average of probabilities for each state. {In both cases, the average is taken for all states in the range $T \in [0, 100] \ K$ after the protocol application.}

The application of the protocol state agrees with the theoretical results for the noiseless simulation, taking all the states to the ground state $\ket{00}$. For the noisy case, the protocol still shows a probability higher than $80\%$ for the ground state, which shows the low influence of {the noise in the quantum information processing, even with the decrease in fidelity at room temperature shown in Fig. \ref{fig:results}}. Thus, simulation reveals that the optimal extraction protocol works with accuracy up to $T<100 \ K$, although the pure state regime discussed in \cite{cruz2022quantum} is limited to $T < 83 K$. 

\begin{figure}
    \centering
    \includegraphics[width=1\linewidth]{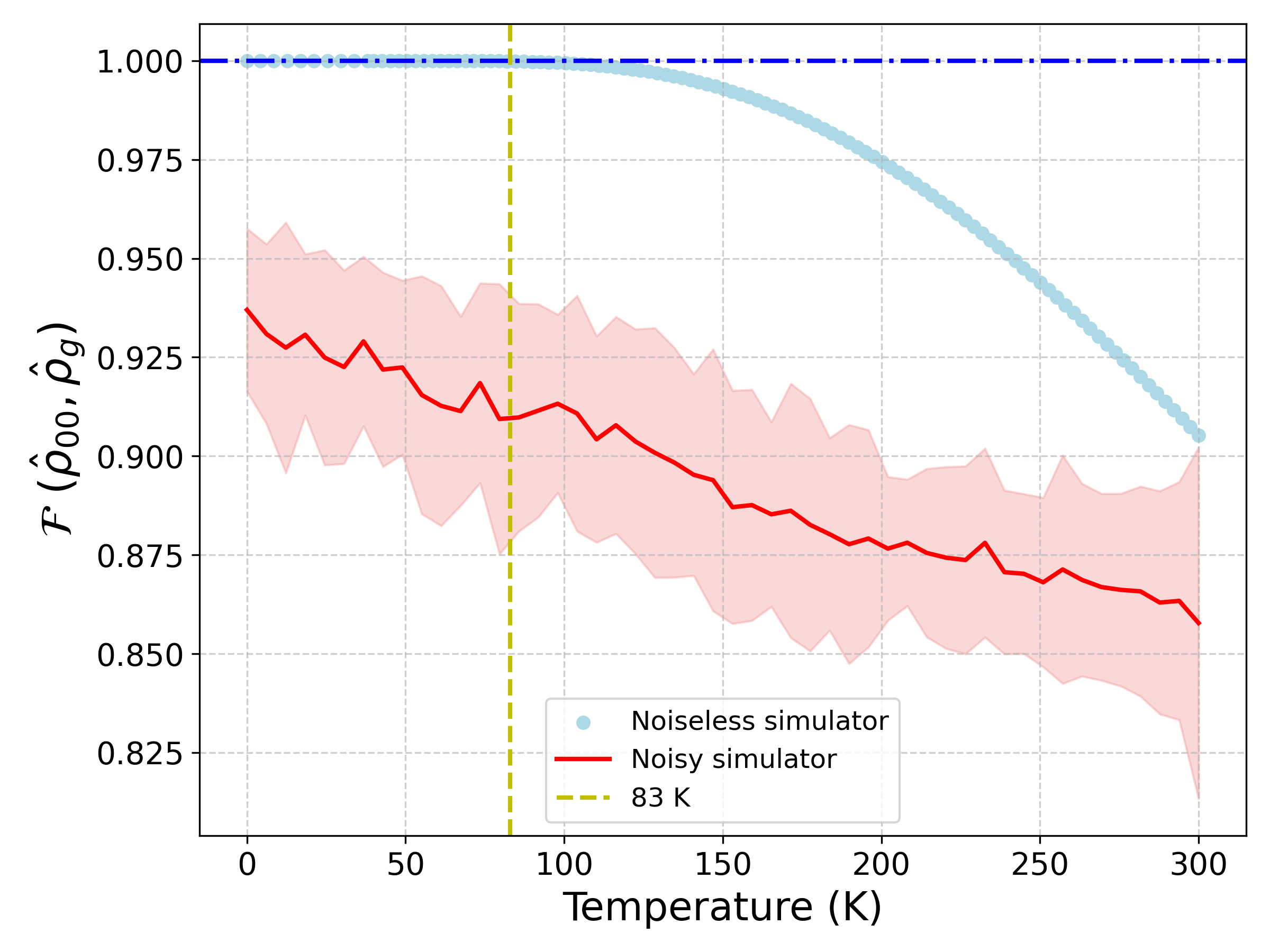}
    \caption{\justifying Fidelity between the simulated state $\rho_{00}$ and the ground state {$\rho_{g}$}. This metric allows visualizing the protocol efficiency in the extraction of the energy taking a maximally entangled to a ground state configuration. The result shows that the optimal extraction protocol can work with significant accuracy for temperatures up to $100 \ K$. The noisy simulation was performed 30 times, with a solid line as its average and the shadow as the standard deviation.} 
    \label{fig:Fid}
\end{figure}

A way to validate the effectiveness of the protocol in the simulation is to see how similar the final state is from the ground state, represented as $\rho_{g} = \ket{00}\bra{00}$. Considering this aspect, we compare both using fidelity to quantify the distinguishability (i.e., the degree of similarity) between the state resulting from an optimal work extraction and the output of the protocol. Since $\rho_{00}$ is a pure state, we use { $\mathcal{F} (\hat \rho_{00}, \hat \rho_g) = \sqrt{\bra{0} \hat \rho_{00}\ket{0}}$} for all temperature points. The results can be seen in Figure \ref{fig:Fid}. As expected, {$\mathcal{F}  (\hat \rho_{00}, \hat \rho_g)$} decrease for $T>100 \ K$ since the ergotropy decrease in that range. 

\begin{figure*}[t!]
 \centering
  \subfloat{\includegraphics[width=.5\linewidth]{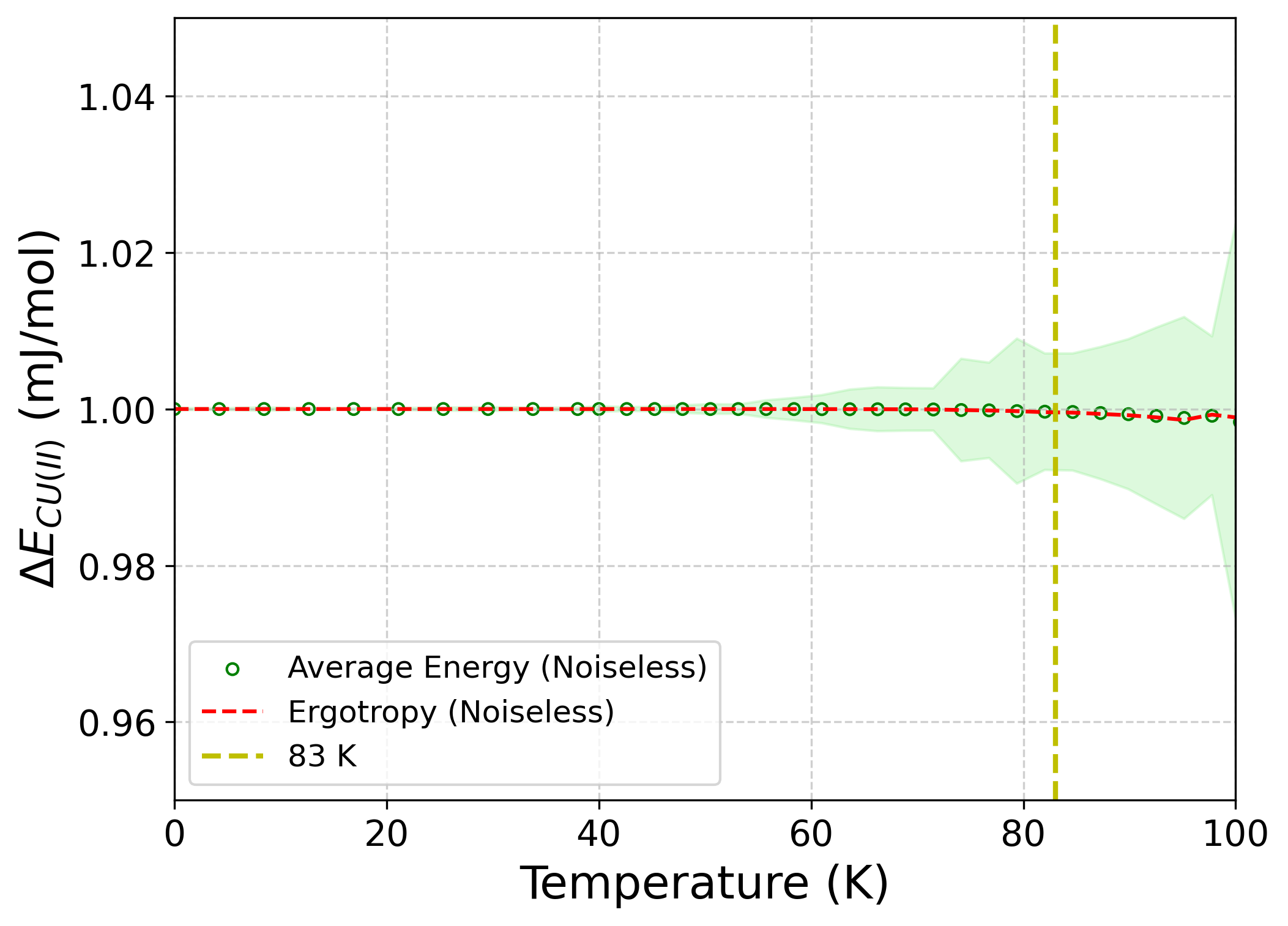}}
  \subfloat{\includegraphics[width=.5\linewidth]{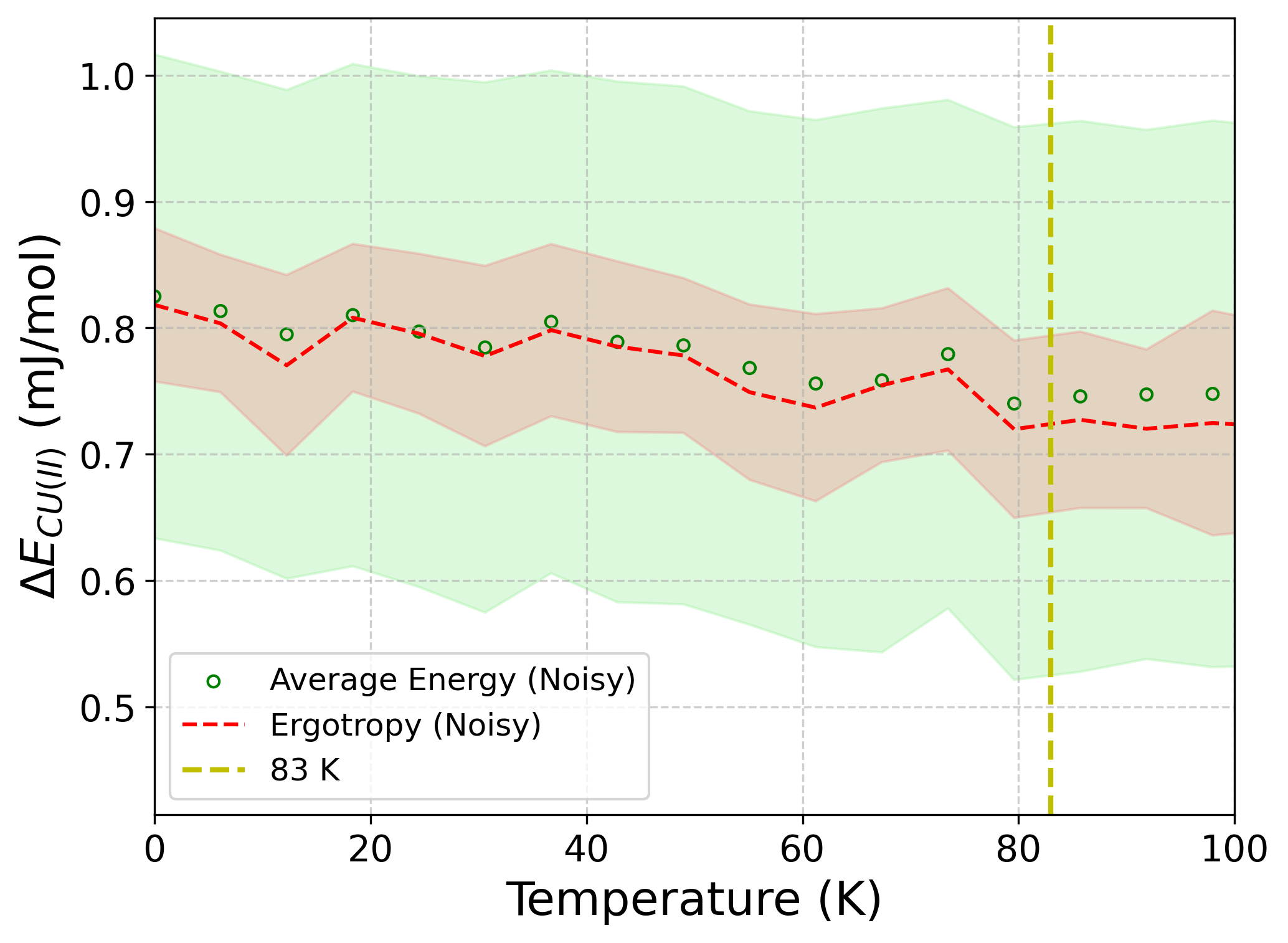}}
  \caption{\justifying {Comparative between average energy and ergotropy for the noiseless (left) and noisy simulation (right). The result shows the protocol works with good accuracy for $T<100 \ K$, extracting the maximum amount of energy stored in the battery for the noiseless scenario. Conversely, the noisy simulation was performed 30 times, with the average represented by the solid line for the ergotropy and the open points for the average energy, while the shadows represent the precision for the average energy.}}
  \label{fig:fluctuations}
\end{figure*}

The principle that not all the energy in the system can be extracted in the form of work is widely known. Using Eq.(\ref{ergotropy}), we can show that our dinuclear battery simulation holds this principle for ergotropy. Another important fact is that different temperatures result in different precisions in the extraction process since the optimal control gets harder for higher temperatures. The precision can be compute considering Eq.(\ref{variance}) and both values can be seen in Fig. \ref{fig:fluctuations} for the noiseless e noisy case. 

{In the noisy case, although the average energy has a close correspondence to ergotropy, there is a striking difference in the precision (shadows) --- the average energy variance denotes a higher impression.} This suggests that noise dissipates energy but maintains a relatively stable fraction of the useful energy extractable as work. Additionally, noise may prevent the system from fully thermalizing, preserving some structure that favors work extraction. Thus, although Ref.\cite{cruz2022quantum} highlights that the system is capable of energy storage at room temperature, our results reveal a high precision at low temperatures, while the inherent imprecision in the extraction process at higher temperatures could compromise the practical application of these devices. On the other hand, despite the overall degradation of energy efficiency, there is still potential for useful energy conversion even in a noisy environment. This is an important result to characterize the efficiency of the battery even in current quantum hardware standards.

\section{Conclusion}\label{sec:conclusion}

In summary, this work demonstrates the potential of VQAs to simulate the work extraction in the practical scenario of low-dimensional metal complexes. The results show that the proposed VQA approach can successfully capture the relationship between ergotropy and the magnetic susceptibility of the compound, showing reasonable agreement with the experimental data for dinuclear copper (II) complexes and capturing key features of the stored energy.  %Using the Variational Quantum Thermalizer algorithm, we modeled the quantum properties and optimal work extraction process in a dinuclear quantum battery, exploring the relationship between the energy extraction and the noisy environment. The simulated noiseless results showed reasonable agreement with the experimental data for dinuclear copper (II) complexes, capturing key features of the stored energy.  

{Although the experimental results align more closely with the noiseless simulation, the small difference between the noisy and noiseless simulations suggests that the impact of noise in the experiment is minimal for ergotropy, but increase for average energy precision.} The results show that the presence of environmental noise hinders the accurate evaluation of the energy stored and retrievable from the battery. Moreover, temperature is another critical limitation. While the recent literature underscores the capabilities of the dinuclear battery model based on metal complexes to store energy at room temperature, the work extraction protocol was found to be highly precise only at low temperatures. As the temperature increases, the quantum correlations that enable efficient work extraction are naturally weakened, leading to a noticeable increase in work fluctuation. In fact, in a noiseless scenario, the simulation of the work extraction protocol exhibits high precision only at lower temperatures,  with performance degradation observed close to $100 \ K$.  Conversely, the precision of the work extraction in a noisy environment reveals that fluctuations in quantum correlations can lead to significant deviations, impacting the efficiency of the protocol. %This underscores the necessity of robust error mitigation strategies to enhance the reliability of this quantum battery model in a real scenario.
These limitations emphasize that, under current conditions, quantum batteries based on dinuclear metal complexes would require careful control of environmental factors or protocol corrections to function reliably.

Therefore, future research should focus on refining the VQA framework to mitigate noise effects and improve its resilience on current quantum hardware. By addressing these challenges, this approach can contribute to the development of viable quantum battery technologies, bridging the gap between theoretical models and experimental implementation. The combination of variational algorithms and quantum thermodynamic models represents an interesting pathway toward realizing the potential of quantum devices on noisy quantum platforms while also guiding the development of improved strategies to overcome current limitations.

\begin{acknowledgments}
The authors thank Fundação de Amparo à Pesquisa do Estado da Bahia - FAPESB for its financial support (grant numbers APP0041/2023 and PPP0006/2024). \\
\end{acknowledgments}
\appendix

\section{Noisy Model}\label{app:noise}

The noise model consists of a well-defined set of basis gates, which include \textit{cx}, \textit{delay}, \textit{id}, \textit{measure}, \textit{reset}, \textit{rz}, \textit{sx}, and \textit{x}. These gates form the fundamental operations upon which the quantum circuits are constructed, ensuring compatibility with the noise model. Among these, specific operations are directly affected by noise, namely \textit{sx}, \textit{id}, \textit{x}, \textit{cx}, and \textit{measure}. The inclusion of noise in these instructions impplies that the model accounts for common decoherence mechanisms, such as relaxation and dephasing, as well as errors arising from gate imperfections.

The noise model explicitly considers two qubits, indexed as 0 and 1, meaning that any noise effects are applied specifically to these qubits rather than a generalization over an arbitrary number of qubits. Each gate or operation subject to noise is associated with a specific qubit or qubit pair, depending on the nature of the gate. For instance, two-qubit errors are modeled for the \textit{cx} gate when applied between qubits (0, 1), while single-qubit errors are introduced for the \textit{id}, \textit{sx}, \textit{x}, and \textit{measure} operations on both qubits.

In particular, the presence of noise in the \textit{id} gate indicates that even idle qubits are affected by decoherence, primarily through $T_1$ relaxation and $T_2$ dephasing. The relaxation time, $T_1$, represents the characteristic time scale over which an excited qubit decays to the ground state due to energy dissipation, while the coherence time, $T_2$, describes the time scale over which phase coherence is lost due to environmental interactions and other stochastic processes. In this model, these values are set as $T_1 = 0.00015774397097652505 \ s$ and $T_2=0.00010861203881817735 \ s$, respectively. The system operates at a frequency of approximately 5.23 GHz, which is consistent with typical superconducting qubit platforms.

Furthermore, the noise model considers the impact of measurement errors, which are applied independently to qubits 0 and 1. Measurement noise accounts for inaccuracies in state readout, which can arise from various sources, including thermal noise in the measurement electronics and imperfections in the quantum-to-classical conversion process. These effects are crucial for realistic simulations, as they introduce an additional layer of uncertainty in the final results.

Table \ref{tab:noise} summarizes the key attributes of the noise model, listing the basis gates, noisy instructions, affected qubits, and specific error sources associated with each operation. The inclusion of these parameters ensures a comprehensive representation of noise effects in the quantum system, enabling more accurate and realistic simulations of quantum circuits under decoherence and gate imperfections.

\begin{table}[h]
\centering
\renewcommand{\arraystretch}{2} % Ajusta o espaçamento entre linhas
\setlength{\tabcolsep}{5pt} % Ajusta o espaçamento entre colunas
%\scriptsize % Reduz o tamanho da fonte
\begin{tabular}{@{}p{2.5cm}p{5.5cm}@{}}
\toprule
\textbf{Attribute}              & \textbf{Configuration}                                                                                   \\ \midrule
{Basis gates}            & \texttt{['cx', 'delay', 'id', 'measure', 'reset', 'rz', 'sx', 'x']}                                     \\ 
{Instructions with noise} & \texttt{['sx', 'id', 'x', 'cx', 'measure']}                                                             \\ 
{Qubits with noise}      & \texttt{[0, 1]}                                                                                         \\ 

{$T_1 (s)$}      & \texttt{0.00015774397097652505}                                                                                         \\ 

{$T_2(s)$}      & \texttt{0.00010861203881817735}  \\

{Frequency(Hz)}      & \texttt{5227644738.696302}  \\

{Specific qubit errors}  & \texttt{[('cx', (0, 1)), ('id', (0,)), ('id', (1,)), ('sx', (0,)), ('sx', (1,)), ('x', (0,)), ('x', (1,)), ('measure', (0,)), ('measure', (1,))]} \\ \bottomrule
\end{tabular}
\caption{ \justifying  Description of the noise model used in the simulation.}
\label{tab:noise}
\end{table}
\newpage
%\bibliography{bib.bib}


\end{document}